\documentclass[final,5p,times,twocolumn]{elsarticle}

\usepackage{amssymb}
\usepackage{amsmath}
\usepackage{siunitx}
\DeclareSIUnit{\sample}{S}
\DeclareSIUnit{\pixel}{px}
\DeclareSIUnit{\ADCcount}{ADC\,counts}
\usepackage[utf8]{inputenc}
\usepackage[T1]{fontenc}
\usepackage{hyperref}
\hypersetup{
    colorlinks=true,
    linkcolor=blue}

\usepackage{array,tabularx,caption}    
\journal{NIM A}

\begin{document}
\begin{frontmatter}

%% Title, authors and addresses

%% use the tnoteref command within \title for footnotes;
%% use the tnotetext command for theassociated footnote;
%% use the fnref command within \author or \address for footnotes;
%% use the fntext command for theassociated footnote;
%% use the corref command within \author for corresponding author footnotes;
%% use the cortext command for theassociated footnote;
%% use the ead command for the email address,
%% and the form \ead[url] for the home page:
%% \title{Title\tnoteref{label1}}
%% \tnotetext[label1]{}
%% \author{Name\corref{cor1}\fnref{label2}}
%% \ead{email address}
%% \ead[url]{home page}
%% \fntext[label2]{}
%% \cortext[cor1]{}
%% \affiliation{organization={},
%%             addressline={},
%%             city={},
%%             postcode={},
%%             state={},
%%             country={}}
%% \fntext[label3]{}

\title{Optical readout of MicroPattern Gaseous Detectors with SiPMs}

\author[a]{F.M.~Brunbauer\corref{cor1}}
\cortext[cor1]{Corresponding author.}
\ead{florian.brunbauer@cern.ch}
%\author[f]{E.~Ferrer-Ribas}
\author[a]{K.J.~Flöthner}
\author[b]{F.~Garcia}
\author[c,d]{D.~Janssens}
\author[a]{E.~Jelinkova}
\author[a]{M.~Lisowska}
\author[a]{D.~Marques}
\author[a,e]{H.~Müller}
\author[a]{E.~Oliveri}
%\author[f]{T.~Papaevangelou}
\author[f]{E.~Pollacco}
\author[a]{L.~Ropelewski}
\author[a]{L.~Scharenberg}
\author[a]{B.~Schönfelder}
\author[a]{E.~Tuovinen}
\author[a]{M.~van Stenis}
\author[a]{R.~Veenhof}

\address[a]{European Organization for Nuclear Research (CERN), CH-1211 Gen\`{e}ve 23, Switzerland}

\address[b]{Helsinki Institute of Physics, University of Helsinki, Finland}

\address[c]{Inter-University Institute for High Energies (IIHE), Universit\'{e} libre de Bruxelles, Boulevard du Triomphe, CP 230, 1050 Brussels, Belgium}

\address[d]{Fonds de la Recherche Scientifique -- FNRS, Rue d'Egmont 5, 1000 Brussels, Belgium}

\address[e]{University of Bonn, Regina-Pacis-Weg 3, 53113 Bonn, Germany}

\address[f]{Institut de Recherche sur les lois Fondamentales de l'Univers (IRFU, CEA), Universit\'{e} Paris Saclay, F-91191 Gif-sur-Yvette, France}

\begin{abstract}

Scintillation light readout of MicroPattern Gaseous Detectors (MPGDs) is used for imaging and track reconstruction applications including in optical Time Projection Chambers (TPC). Typically, high resolution CCD or CMOS imaging sensors are used and PMTs can be employed to obtain auxiliary timing information. We report on the optical readout of MPGDs with Silicon Photomultipliers (SiPMs) which enables limited spatial resolution. Arrays of SiPMs were used to read out Gaseous Electron Multipliers (GEMs) and Glass Micromegas amplification stages. Center of gravity reconstruction of hit positions yields a spatial resolution of $\approx\SI{1.35}{\milli\metre}$ for signals shared across multiple SiPMs. High gain of the amplification stage as well as the SiPMs enables good signal sharing and is advantageous for position reconstruction.
%SiPMs can also be used to reconstruct particle tracks in optical TPCs where the arrival time information encoded in the SiPM waveform can be used to determine the relative depth of interaction along the particle trajectory. We report not on the simultaneous optical readout of a TPC with SiPMs and a CCD camera to combine the high spatial resolution of the imaging sensor with timing information from the SiPM.

\end{abstract}


\end{frontmatter}

%\begin{linenumbers}

%\linenumbers

%% main text
%\section{Sample Section Title}
%\label{sec:sample1}

%% For citations use: 
%%       \citet{<label>} ==> Jones et al. [21]
%%       \citep{<label>} ==> [21]
%%

\section{Introduction}
\label{sec:sample1}

%\begin{linenumbers}
MicroPattern Gaseous Detectors (MPGDs) combine charge amplification in microscopic electric-field structures with fine spatial granularity, high rate capability and a low material budget. Prominent examples include Gaseous Electron Multipliers (GEMs) \cite{Sauli1997GEM}, in which electron multiplication occurs inside holes in a metal-clad insulating foil, and MicroMesh Gaseous Structures (Micromegas) \cite{Giomataris1996Micromegas}, which employ a narrow amplification gap between a micromesh and an anode. These structures amplify the primary ionisation produced by incident radiation and provide a versatile basis for particle tracking and radiation imaging.

The recording of scintillation light emitted during electron avalanche multiplications in MPGDs has been demonstrated to achieve high resolution with state-of-the-art imaging sensors \cite{Fraga2001Optical, glassMicromegas}. In combination with fast photon detectors such as Photomultiplier Tubes (PMTs), optical readout can be used for 3D track reconstruction in Time Projection Chambers where 2D information from imaging sensors is combined with depth information from PMTs. This concept is employed for directional dark matter searches by the CYGNO \cite{cygnoExperiment} experiment as well as for the search for the Migdal effect \cite{migdalExperiment}. Optical readout relies on a good matching of the emission spectrum of the gas with the wavelength-dependent sensitivity of the recording device. The visible light sensitivity of most imaging sensors has led to the popularity of CF$_{4}$ as scintillation gas as it features a strong visible scintillation band peaked around \SI{630}{\nano\metre} \cite{cf4GEMScintillation}. While CF$_{4}$ scintillation is well-matched to most imaging sensors, it presents a strong constraint on the gas choice and it is also facing decreasing availability due to its classification as strong greenhouse gas \cite{EU_FGas_2024}.

Scintillation light readout offers considerable flexibility in the choice and arrangement of the sensing elements. CCD and CMOS cameras provide finely sampled two-dimensional images, while optical magnification allows the field of view and effective sampling at the amplification plane to be adjusted. The optical connection electrically separates the photosensor from the amplification electrodes and allows high granularity without a correspondingly segmented charge-collection anode. Complementary fast photodetectors can measure the arrival-time distribution of the scintillation signal, providing information on electron drift and enabling three-dimensional reconstruction in time projection chambers \cite{Bihaowicz2014,BRUNBAUER201824}.
Optical readout has been demonstrated with multiple MPGD technologies, most notably Gaseous Electron Multipliers and Micromegas. While the open geometry of GEMs and their use in multi-GEM stacks is ideally suited for optical readout by imaging sensors placed below the last GEM, Micromegas had to be implemented on a transparent substrate to allow recording of emitted scintillation light \cite{glassMicromegas}. Glass Micromegas have been used for X-ray and beta imaging as well as for neutron detection with good spatial resolution \cite{Cools2023a}.
Silicon photomultipliers (SiPMs) offer a compact approach to combining spatially resolved optical detection with fast timing. Precise timing readout of a gaseous detector has been demonstrated with a combination of PICOSEC Micromegas with SiPM readout \cite{brunbauer2026sipmopticalreadoutpicosec}. Comprising many avalanche photodiode microcells operated in Geiger mode, SiPMs provide single-photon sensitivity, high internal gain and a fast response. Arrays of independently read out SiPMs can measure the distribution of scintillation light across the sensor plane.

In this work, we report on the readout of MPGDs with SiPMs which provide spatial resolution depending on the readout configuration. 

%from arrays of photosensors with the good timing resolution of these devices. This enables the determination of particle hit locations as well as the reconstruction of particle tracks in TPC detectors. 

Combined with the high resolution of camera, SiPMs could potentially help with 3D track reconstruction in a similar way as PMTs while their small size allows multiple SiPMs to be placed in an array after the amplification stage. 
%This is advantageous for non-straight particle trajectories as timing information is obtained from multiple angles.

\section{Experimental methodology}
\label{sec:3}

\begin{figure}[!t]
\begin{center}
\includegraphics[width=\columnwidth]{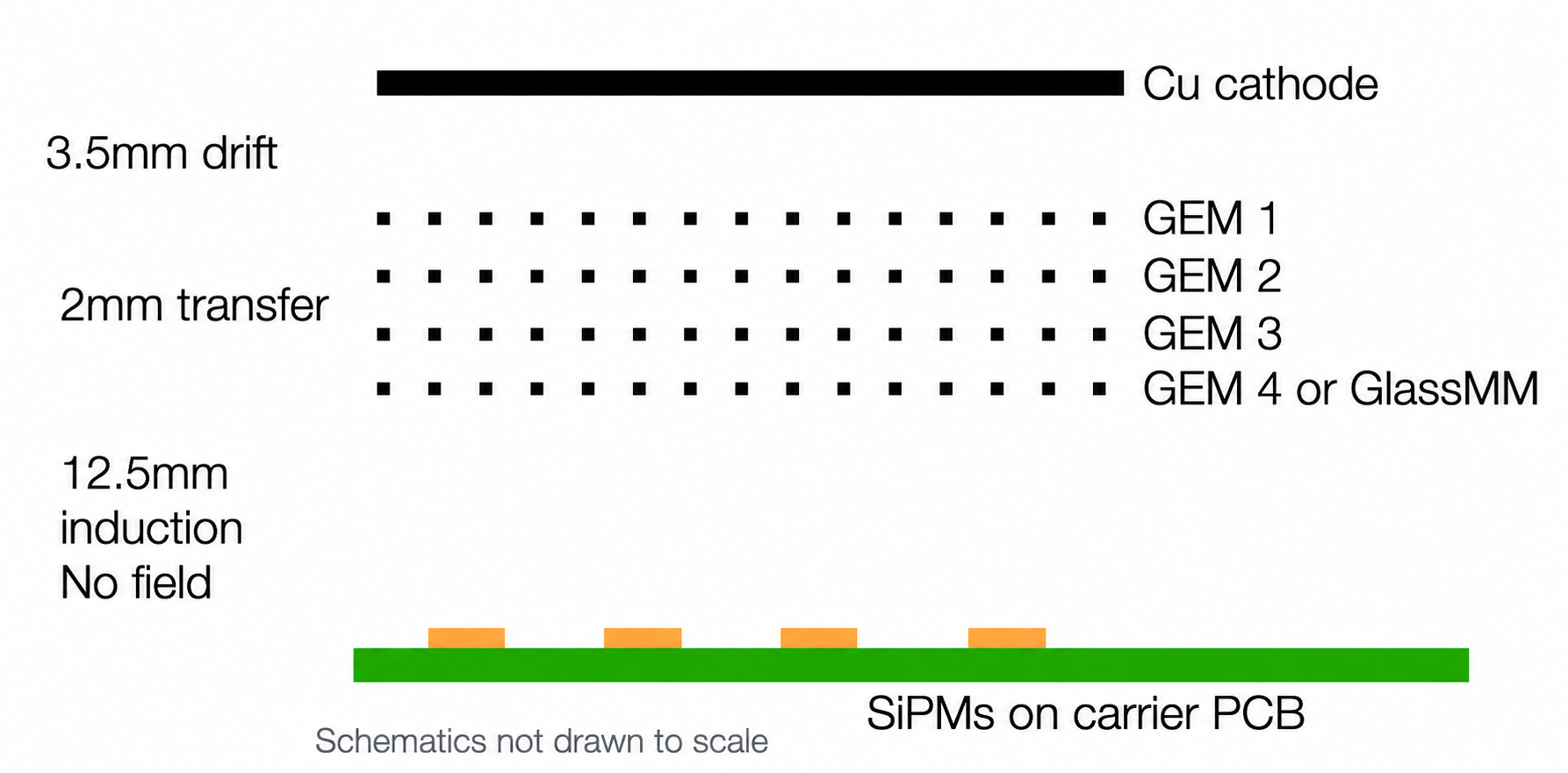}
\end{center}
\caption{Crosssection of detector setup showing stack of 3-4 GEMs and placement of SiPMs below the amplification structure.}
\label{schematicCrosssection}
\end{figure}

A quad-GEM detector was read out with an array of 16 SiPMs arranged in a 4x4 matrix. A spacing of either \SI{2}{\centi\metre}, \SI{1}{\centi\metre} or \SI{0.5}{\centi\metre} between SiPMs was used in both directions. The SiPMs were placed \SI{12.5}{\milli\metre} below the surface of the last GEM in the stack as shown in figure \ref{schematicCrosssection}. 

\begin{figure}[!t]
\begin{center}
\includegraphics[width=\columnwidth]{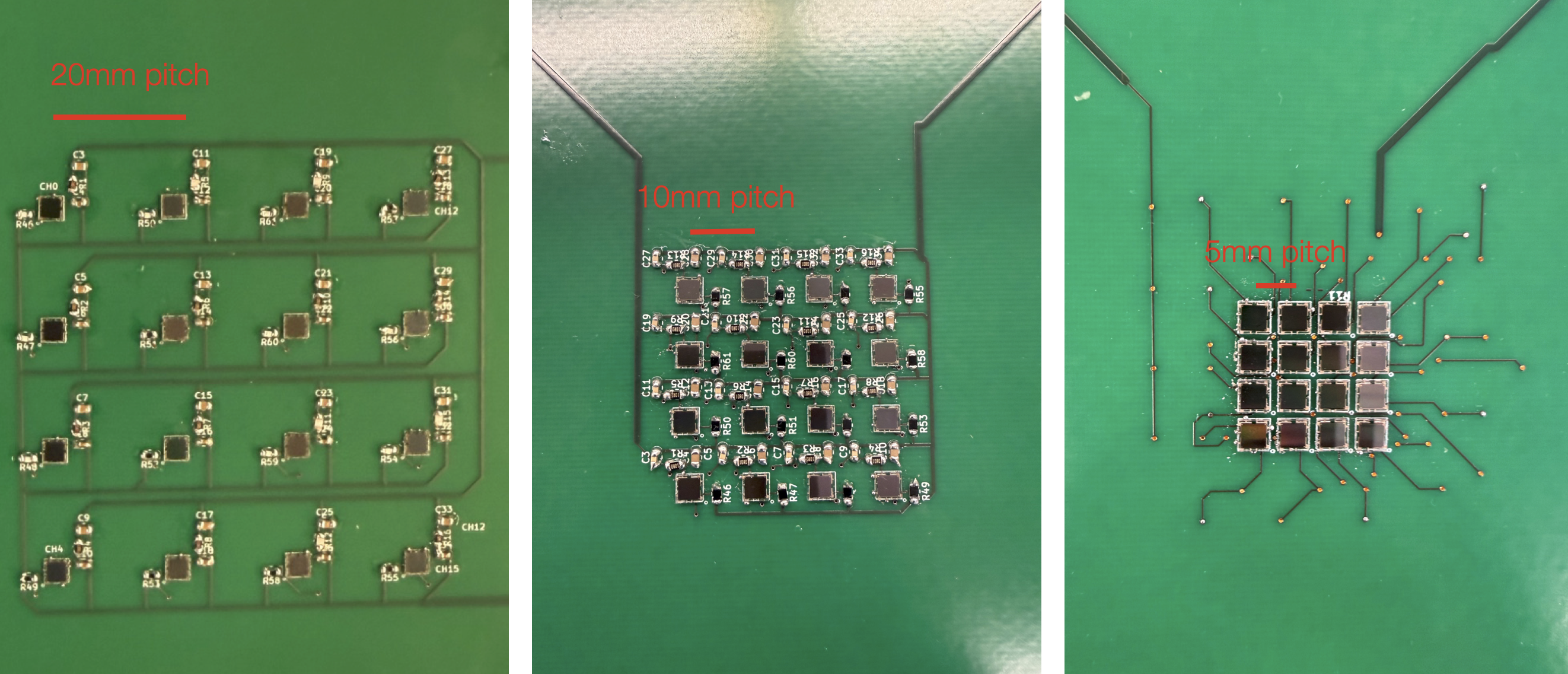}
\end{center}
\caption{Three different SiPM arrays with pitch varying from \SI{20}{\milli\metre} to \SI{10}{\milli\metre} and \SI{5}{\milli\metre}.}
\label{sipmArrays}
\end{figure}

The SiPM readout plane with different pitch between the individual channels is shown in figure \ref{sipmArrays}.

\begin{figure}[!t]
\begin{center}
\includegraphics[width=\columnwidth]{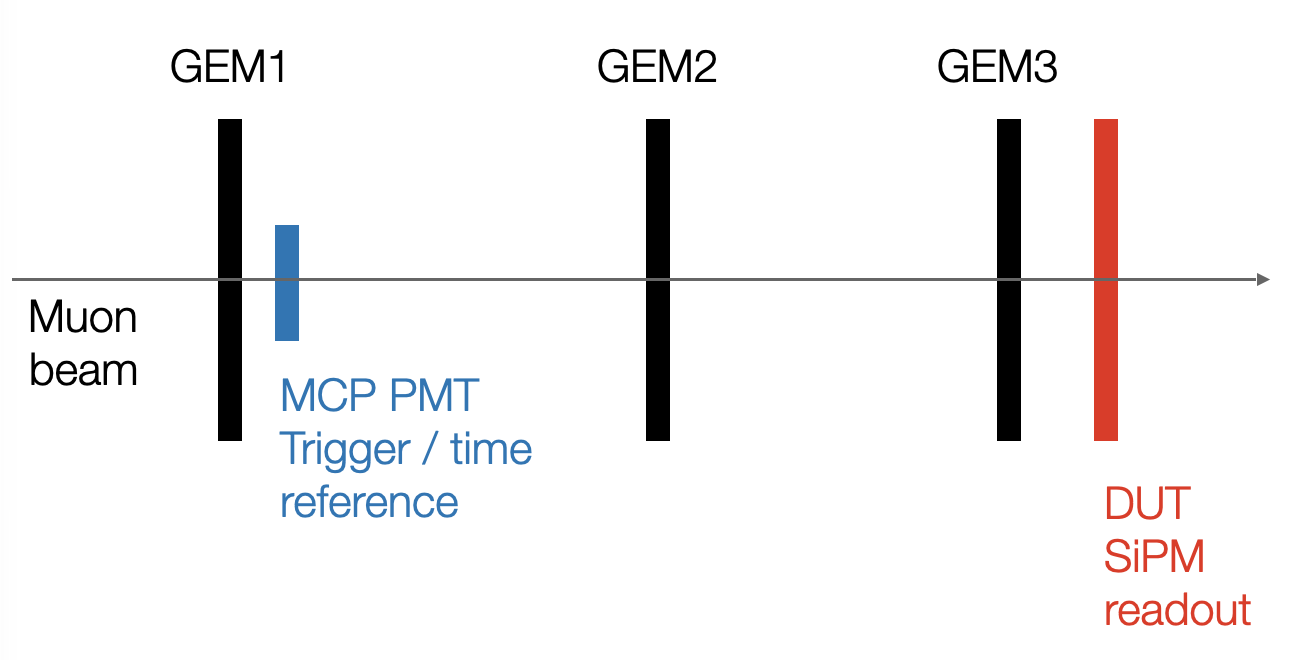}
\end{center}
\caption{Schematic of test beam setup showing the GEM-based tracking telescope, the MCP-PMT used as trigger and time reference and the placement of the SiPM detector.}
\label{testbeamSetup}
\end{figure}

Standard GEMs with a size of $\SI{10}{\centi\metre} \times \SI{10}{\centi\metre}$ were used with a Kapton thickness of \SI{50}{\micro\metre}, holes with a diameter of \SI{70}{\micro\metre} and a hole pitch of \SI{140}{\micro\metre}. GEMs were placed with \SI{2}{\milli\metre} transfer gaps between them and a \SI{3.5}{\milli\metre} thick drift region was defined between the top GEM and a Cu plane cathode. The detector elements including the SiPMs were situated in a gas volume flushed with Ar/CF$_{4}$ (80/20 \%) gas mixture at a rate of \SI[per-mode=symbol]{5}{\litre\per\hour} at atmospheric pressure. SiPMs with an active area of $\SI{3}{\milli\metre} \times \SI{3}{\milli\metre}$ from OnSemi were used (brand: onsemi, type: MICROFC-30035-SMT-TR, \SI{35}{\micro\metre} microcell size \cite{onsemi_c_series}). All SiPMs were powered with a common bias voltage but with separate capacitors close to each device. Each SiPM was separately read out by a CAEN digitiser with a sampling frequency of \SI[per-mode=symbol]{500}{\mega\sample\per\second}.
GEMs were powered with a tapered resistive voltage divider with a voltage of \SI{390}{\volt} across the last GEM. Transfer fields were set to \SI[per-mode=symbol]{650}{\volt\per\centi\metre} and the drift field was set to approximately \SI[per-mode=symbol]{330}{\volt\per\centi\metre}.
The detector was operated in a muon test beam at the SPS H4 beam line. A muon beam with an energy of \SI{150}{\giga\electronvolt} was used which covered the full area of the detector. A GEM-based tracking telescope was employed to obtain the true hit locations of particles. Tracks were reconstructed from hits in the three planes of the tracking telescope with a resolution of better than \SI{100}{\micro\metre}. The tracking telescope produced an event counter signal, which was recorded by the digitisers used to read out the SiPMs to synchronise the devices. The tracking telescope was triggered with an MCP-PMT providing a trigger signal. A schematic of the test beam setup is shown in figure \ref{testbeamSetup}.

The MCP-PMT had an active area of \SI{1}{\centi\metre} diameter and was scanned across the active area of the detector under test with movable stages. The small acceptance of the MCP-PMT and the dead time of the telescope limited data taking to about 1000 events per spill.

%For TPC readout, a 4x4 array of the same SiPM was used, which was mounted on a PCB with ribs to support the SiPMs and minimise the obstruction of the view between them. Simultaneously, optical readout was performed with a CCD camera mounted outside of the gas vessel and viewing the GEMs through a viewport. A QImaging Retiga R6 CCD camera was used which features \SI{6}{\mega\pixel} resolution and a read noise of XX electrons. The camera was operated in binning mode to increase its signal to noise ratio. Images were acquired with an exposure time of \SI[parse-numbers=false]{XX}{\milli\second}. The camera exposure signal was used to trigger the digitisers to achieve synchronisation between the SiPM readout and the acquired camera frames. Background images were acquired with the detector switched off and averaged background images were subtracted from the camera frames.
%The TPC featured a \SI{10}{\centi\metre} diameter drift region with a length of \SI{10}{\centi\metre}. Field shaper rings placed with a spacing of \SI{1}{\centi\metre} were used to define a uniform drift field along the full length of the active volume between a thin foil cathode and the GEM stack. A drift field of approximately \SI[per-mode=symbol]{100}{\volt\per\centi\metre} was used.

\begin{figure}[!t]
\begin{center}
\includegraphics[width=\columnwidth]{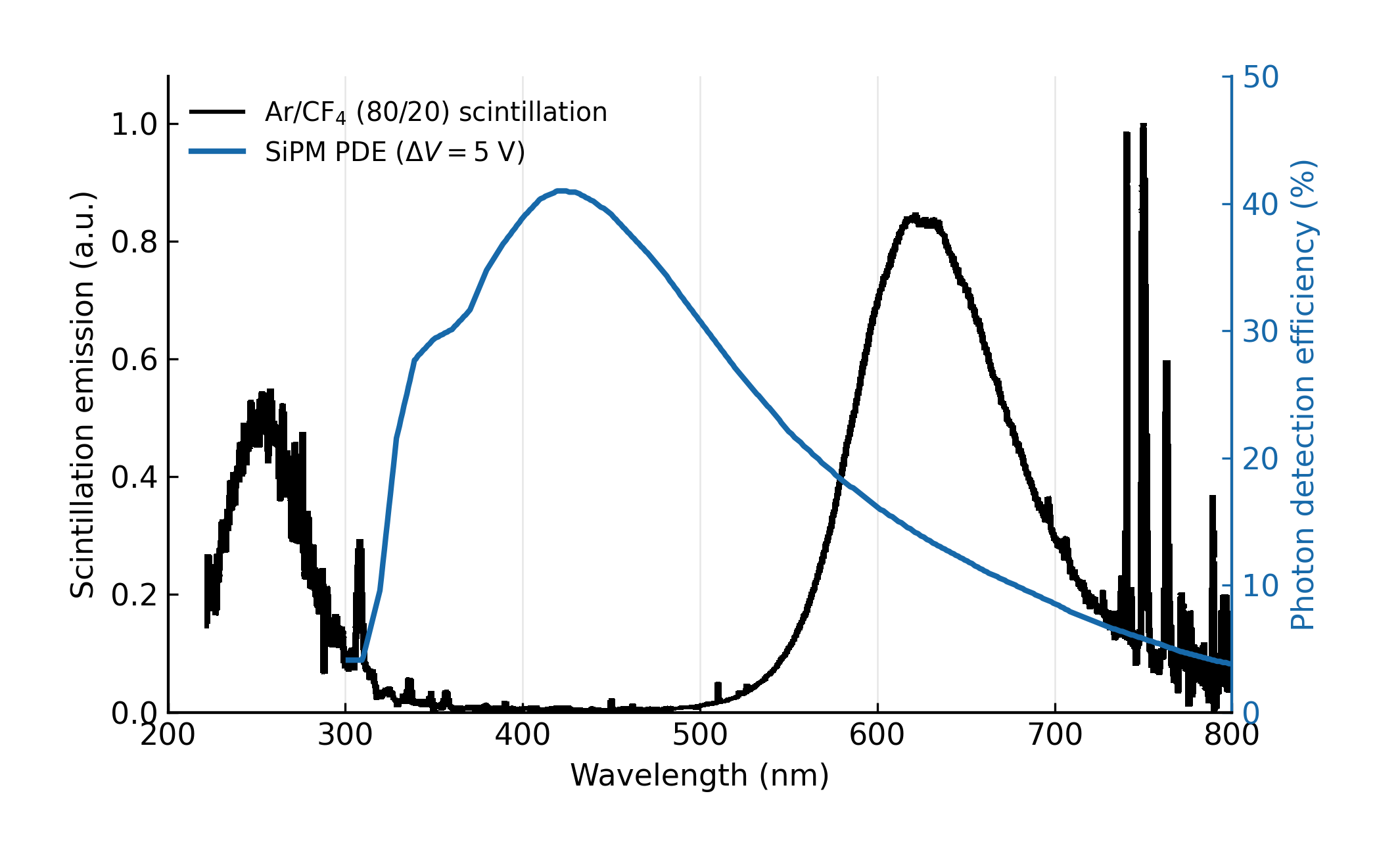}
\end{center}
\caption{Ar/CF$_{4}$ (80/20 \%) secondary scintillation spectrum and wavelength-dependent photon detection efficiency (PDE) of the employed SiPMs for \SI{5}{V} overvoltage from datasheet.}
\label{spectrumSensitivity}
\end{figure}

The scintillation spectrum of CF$_{4}$ is shown in figure \ref{spectrumSensitivity} along with the wavelength-dependent sensitivity of the SiPMs. As shown, the visible scintillation band of CF$_{4}$ is well matched to sensitivity of the readout devices. An alternative to using CF$_{4}$ is the use of wavelength shifters to convert UV light emitted by other gas mixtures to the visible wavelength range. This possibility was explored with a combination of TPB used as solid wavelength shifter coated on the anode of a glass Micromegas detector. TPB converts UV light over a wide wavelength range to a visible band around \SI{450}{\nano\metre} which can be picked up with SiPMs.

\section{Results}
\label{sec:4}

\subsection{SiPM array readout of GEM-based detector}
\label{sec:sipmReadoutGEM}

Particle hit locations were reconstructed with a center of gravity algorithm with a Matlab script. Before reconstruction, several steps of data filtering were applied. A fixed threshold of \SI{30}{\ADCcount} was applied to the SiPM signals to reject noise and dark counts. Signals below this threshold were set to zero and only signals above were considered for position reconstruction. After this threshold, the number of active SiPM channels in each event was extracted. A significant number of events show only one or two SiPM channels above the threshold. Such events are not well suited for position reconstruction as the center of gravity algorithm will reconstruct them exactly at the position of a SiPM or on a line between two adjacent SiPMs. Thus, data is filtered by requiring that at least four SiPMs are active and show a signal above the threshold level defined above. Finally, direct hits of particles on the SiPMs may produce large signals which do not originate from scintillation light emitted by the GEMs but may be produced by direct interaction in the SiPM device. To reject such interactions, a maximum permissible ratio of the amplitude of the highest amplitude SiPM signal and the average of the signals of all other active SiPM was defined and only events with a ratio smaller than 10 were accepted. Finally, the active area is restricted to a range which is fully covered by the SiPM array removing area along the edges where the limited number of SiPMs is insufficient to provide symmetric information about the incident light distribution.
After these filtering steps, the interaction position is reconstructed using a center-of-gravity
(CoG) algorithm applied to the signal amplitudes of the 16 silicon
SiPMs:
\begin{equation}
    \mathbf{r}_{\mathrm{CoG}}
    =
    \frac{\displaystyle\sum_{i=1}^{16} A_i \mathbf{r}_i}
         {\displaystyle\sum_{i=1}^{16} A_i},
    \label{eq:cog_reconstruction}
\end{equation}
where $\mathbf{r}_{\mathrm{CoG}} = (x_{\mathrm{CoG}}, y_{\mathrm{CoG}})$
is the reconstructed position in the sensor plane,
$\mathbf{r}_i = (x_i, y_i)$ denotes the center position of the $i$th
SiPM, and $A_i$ is its measured signal amplitude. The index
$i \in \{1,\ldots,16\}$ runs over all SiPMs.

\begin{figure}[!t]
\begin{center}
\includegraphics[width=\columnwidth]{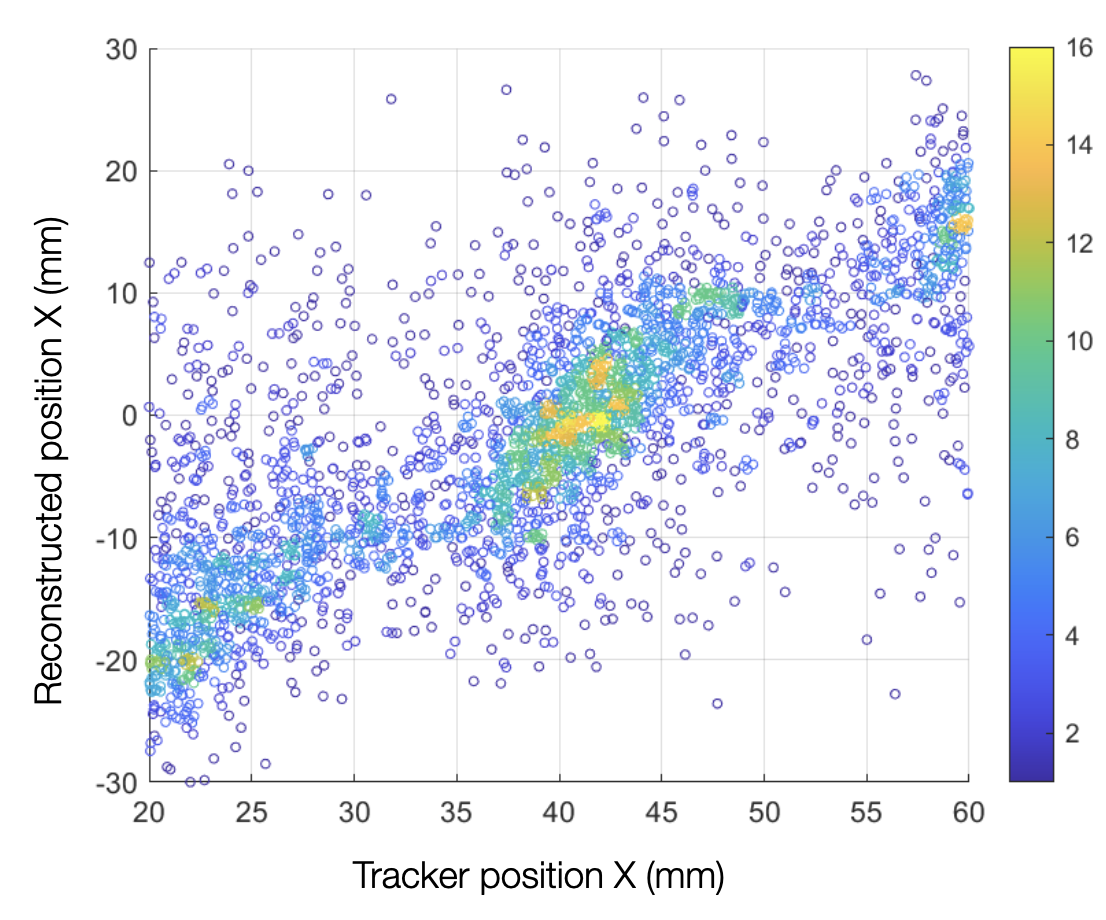}
\end{center}
\caption{Distorter correlation of reconstructed and true X coordinates for the case of \SI{20}{\milli\metre} pitch between SiPMs.}
\label{20mmPitchReconstruction}
\end{figure}

The result should be a linear correlation of the reconstructed position with the true position from the reference tracker. While a good linear correlation of reconstructed position with respect to the reference position was achieved with \SI{5}{\milli\metre} and \SI{10}{\milli\metre} pitch of SiPMs, for the case of \SI{20}{\milli\metre} pitch, insufficient light sharing between neighbouring channels resulted in a distorted correlation as shown in figure \ref{20mmPitchReconstruction}.

%A comparison of the 2D maps of geometrically selected events recorded by the tracker (true position) and obtained by center of gravity reconstruction from the SiPMs is shown in figure XX. 

\begin{figure}[!t]
\begin{center}
\includegraphics[width=\columnwidth]{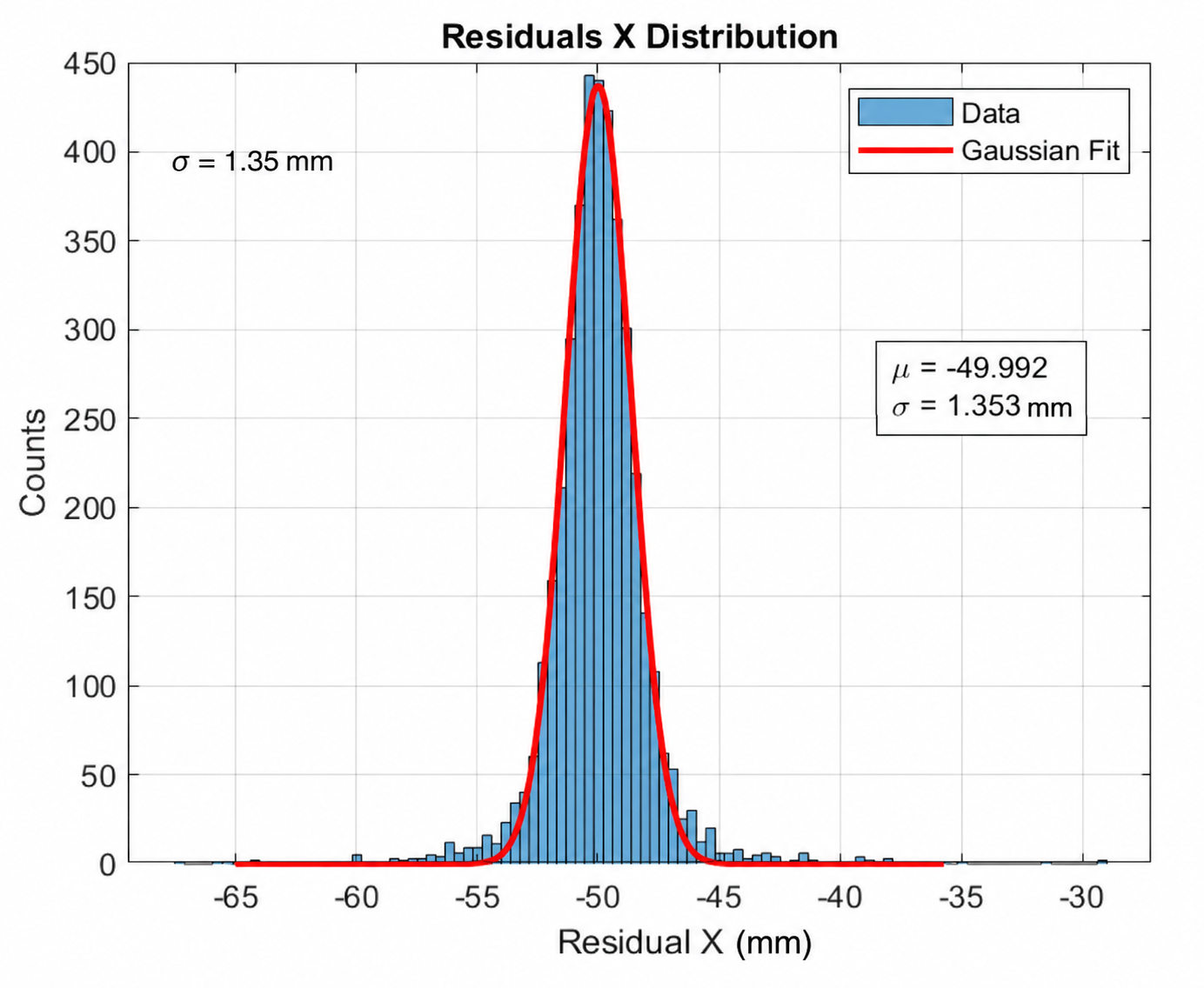}
\end{center}
\caption{X residual distribution for SiPM readout with \SI{10}{\milli\metre} pitch.}
\label{10mmPitchResidualX}
\end{figure}

To quantify the spatial resolution, X and Y residuals defined as the difference between the reconstructed coordinate and the reference coordinate from the tracker were computed. A distribution of the X residuals obtained with a SiPM pitch of \SI{10}{\milli\metre} with a Gaussian fit is shown in figure \ref{10mmPitchResidualX}.

The distribution shows non-Gaussian tails, thus the quoted $\sigma$ refers to the core part of the residuals distribution. A double Gaussian fit may be used to better match the observed tails.

A spatial resolution of $\sigma = \SI{1.35}{\milli\metre}$ was achieved for the configuration with $\SI{10}{\milli\metre}$ pitch while the residuals in the Y coordinate were slightly worse. This is better than the pitch~$/\sqrt{12}$ resolution expected for binary readout, i.e. 2.9, 1.4 and \SI{5.8}{mm} for 10, 5 and \SI{20}{mm} pitch, respectively. However, the achievable resolution appears to be limited by the low number of SiPM channels active for each event. Thus, a smaller distance between SiPM channels should be employed to obtain a higher number of SiPM channels active for each event and improve COG reconstruction capabilities due to increased signal sharing. However, the configuration with $\SI{5}{\milli\metre}$ yielded a worse resolution of $\sigma = \SI{1.7}{\milli\metre}$ and the larger pitch of \SI{20}{mm} configuration only achieved $\sigma = \SI{4.1}{\milli\metre}$. In the case of \SI{5}{mm} pitch, the obtained resolution is worse than the binary readout expectation which could be due to the relatively large size of each SiPM compared to the pitch in that case. The differences between the configurations were also qualitatively reproduced by a ray tracing simulation which showed that the GEM-to-SiPM distance needs to be optimised for each pitch configuration.

\begin{figure}[!t]
\begin{center}
\includegraphics[width=\columnwidth]{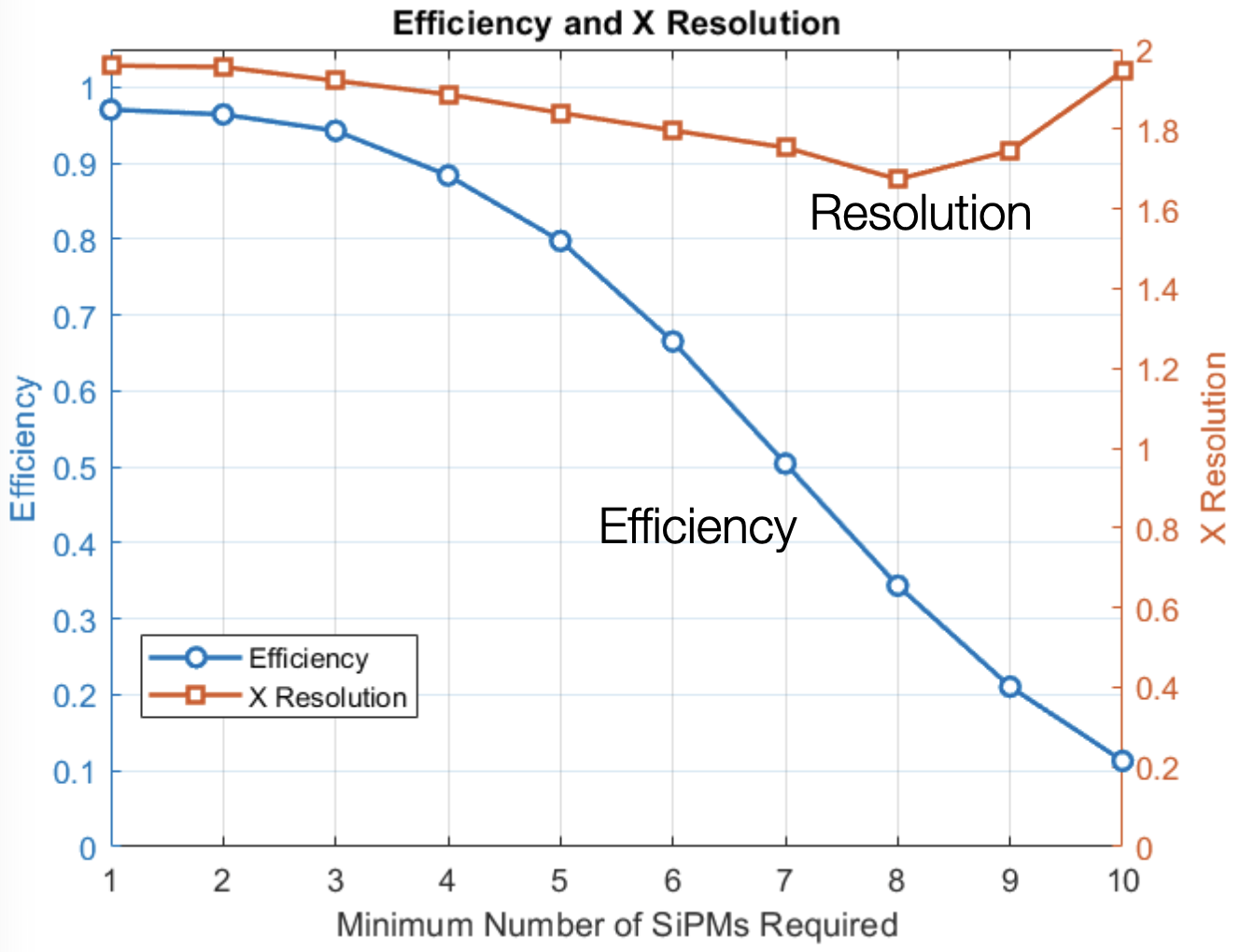}
\end{center}
\caption{Achievable spatial resolution and efficiency as a function of minimum number of SiPMs to accept event for the case of \SI{10}{\milli\metre} pitch between SiPM channels.}
\label{efficiencyResolutionScan}
\end{figure}

The requirement that a minimum of four SiPMs show a signal above a threshold to accept them for position reconstruction has direct implications on the efficiency of the readout. Requiring a higher number of SiPMs to participate in each event would result in a better spatial resolution but lead to inefficiencies due to rejected events. The threshold of four SiPMs required was chosen based on the comparison of achievable spatial resolution and efficiency as a function of the minimum number of SiPMs required shown in figure \ref{efficiencyResolutionScan} where a threshold of four corresponds to an efficiency close to \SI{90}{\percent} of events. The resolution values reported here were obtained with the same CoG reconstruction method described above. However, they used a different spatial selection to probe efficiency over a wider area which may explain the difference in obtained X residuals resolution.

\subsection{SiPM array readout of glass Micromegas detector with TPB wavelength shifter}

\begin{figure}[!t]
\begin{center}
\includegraphics[width=\columnwidth]{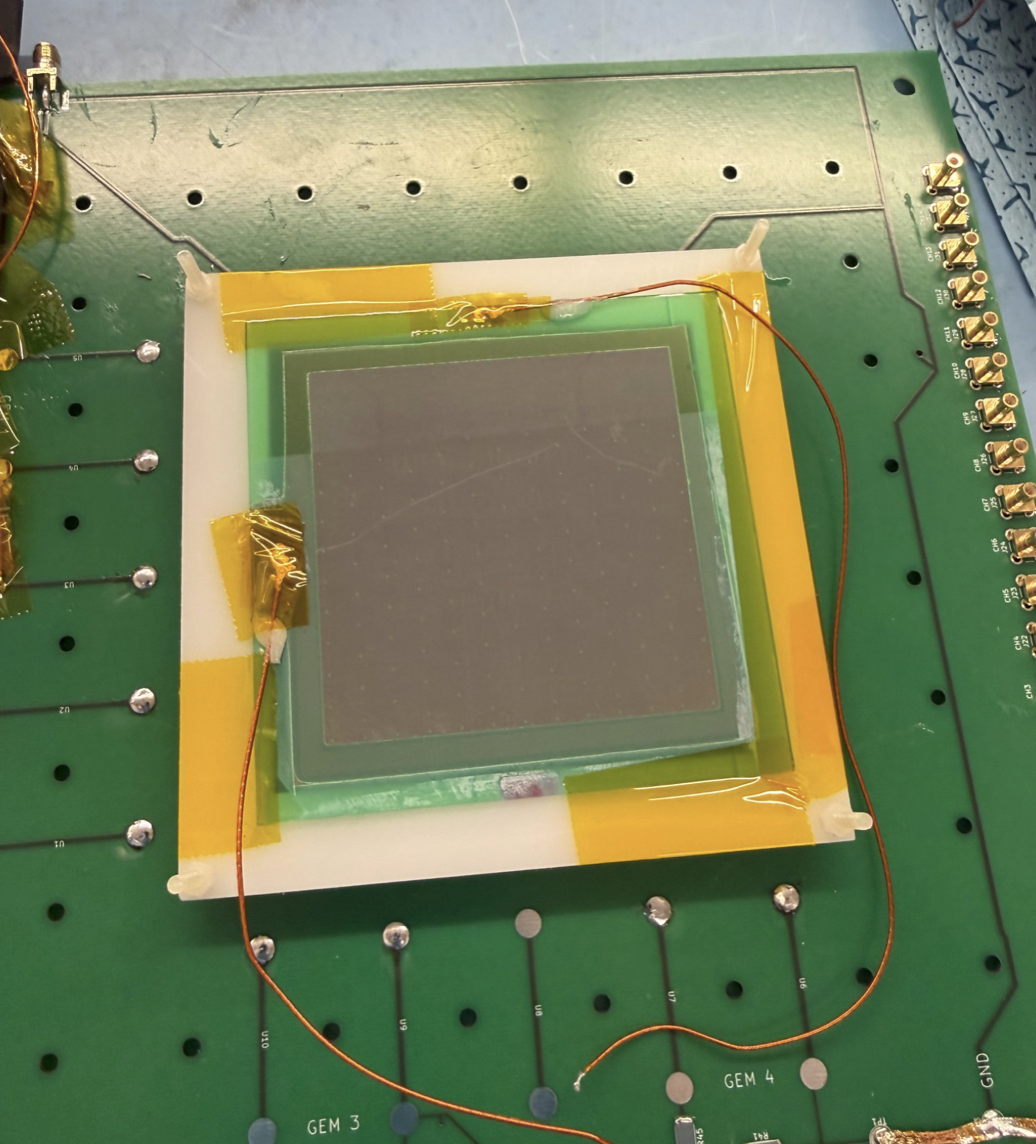}
\end{center}
\caption{Glass Micromegas with TPB coating on anode.}
\label{glassMMTPB}
\end{figure}

A hybrid detector made of a triple GEM and a glass Micromegas was read out with the same SiPM array with a \SI{10}{\milli\metre} pitch. The triple GEM used \SI{2}{mm} transfer gaps and the same transfer gap was used between the GEM and Micromegas. The final amplification stage was a bulk Micromegas with a gap of \SI{128}{\micro m} between the micromegas (\SI{18}{\micro m} diameter wires with \SI{45}{\micro m} openings, woven, calendered) and the anode. The glass Micromegas had a solid wavelength shifting layer of TPB with a thickness of approximately \SI{0.9}{\micro m} deposited on the ITO anode. The ITO anode was a low resistivity coating with a sheet resistivity of 4 $\Omega$ per sq. An image of the TPB-coated glass Micromegas is shown in figure \ref{glassMMTPB}.

This configuration minimised the distance between the scintillation light emission and the conversion to visible light by the TPB layer. This detector was operated in an Ar/CO$_2$ (\SI[parse-numbers=false]{70/30}{\percent}) gas mixture which does not feature significant visible scintillation. The UV scintillation from the mixture is absorbed by the TPB layer and re-emitted at \SI{450}{\nano\metre} which can in turn be recorded by the SiPMs.

The detector was operated with a grounded micromesh and \SI{370}{V} applied on the ITO anode.

\begin{figure}[!t]
\begin{center}
\includegraphics[width=\columnwidth]{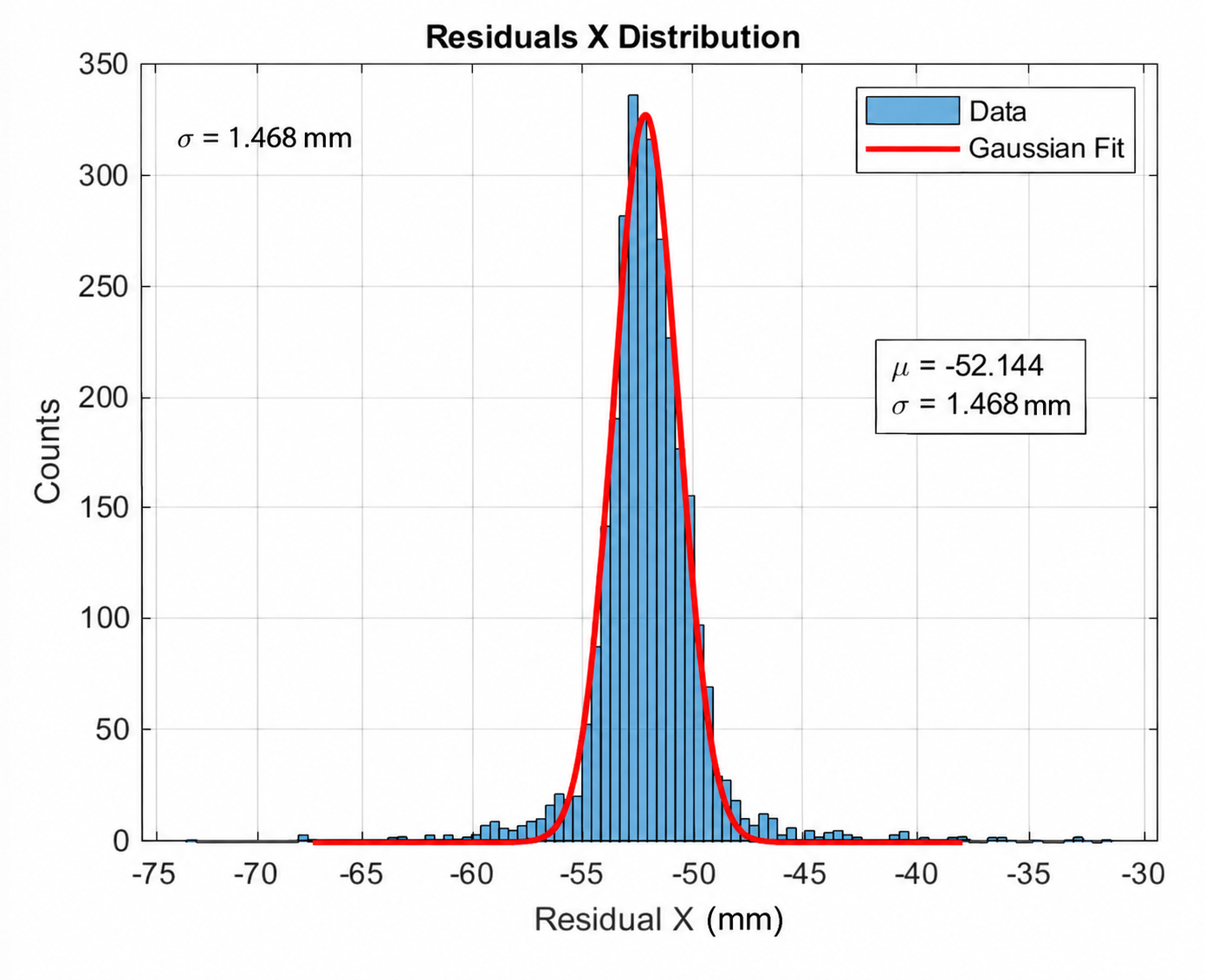}
\end{center}
\caption{X residual distribution for SiPM readout with \SI{10}{\milli\metre} pitch of a TPB-coated glass Micromegas in Ar/CO$_2$ (\SI[parse-numbers=false]{70/30}{\percent}).}
\label{glassMMTPBResidualX}
\end{figure}

A distribution of the residuals in X direction obtained with the glass Micromegas with integrated WLS is shown in figure \ref{glassMMTPBResidualX}. The achieved spatial resolution of $\sigma = \SI{1.46}{\milli\metre}$ is comparable to the value achieved with the \SI{10}{\milli\metre} pitch SiPM array with the quad-GEM detector described above. This demonstrates that similar performance can be achieved in gas mixtures without CF$_{4}$ with the combination of a WLS with SiPM array readout.

\subsection{Machine learning methods for particle reconstruction}

In addition to the CoG reconstruction, different machine learning methods were investigated to determine whether the measured SiPM signals contain additional information that could improve particle-hit position reconstruction. The main objective was to investigate whether the relationship between the measured signals and the particle-hit position could be learned from the measured data, and whether more complex models could capture relationships that are not represented by simpler reconstruction methods.

Four machine learning models were evaluated: Linear Regression, Random Forest, a Multi-Layer Perceptron (MLP), and a one-dimensional Convolutional Neural Network (CNN). Linear Regression, Random Forest, and the MLP used features extracted from the recorded waveforms, including integrated signal, peak amplitude, peak position, and total event signal. The CNN instead used digitised signal waveforms to examine whether the waveform shapes contain additional useful information.

To evaluate the models, the detector area was divided into spatial blocks based on the reference particle-hit positions. Entire blocks were assigned to training, validation, or test sets, allowing the models to be tested on detector regions not included in training. Five spatial splits were evaluated, with all methods compared on the same held-out events within each split. The reconstruction performance was evaluated using the X residual distribution and the mean two-dimensional Euclidean reconstruction error. The results for the three SiPM configurations are summarized in Table~\ref{tab:ml_reconstruction_results}.

\begin{table}[t]
\centering
\caption{Comparison of reconstruction methods for the three SiPM configurations. The best result for each metric and configuration is shown in bold.}
\label{tab:ml_reconstruction_results}
\resizebox{\columnwidth}{!}{%
\begin{tabular}{llcc}
\hline
\textbf{SiPM pitch} & \textbf{Method} &
\textbf{$\sigma_x$ [mm]} &
\textbf{Mean 2D error [mm]} \\
\hline
20 mm & CoG               & \textbf{4.065} & \textbf{8.772} \\
      & Linear Regression & 5.418 & 12.179 \\
      & Random Forest     & 5.503 & 12.865 \\
      & MLP               & 8.608 & 12.699 \\
      & Waveform CNN      & 6.929 & 12.872 \\
\hline
10 mm & CoG               & \textbf{1.134} & 2.415 \\
      & Linear Regression & 1.438 & \textbf{2.046} \\
      & Random Forest     & 2.115 & 3.079 \\
      & MLP               & 2.158 & 2.619 \\
      & Waveform CNN      & 2.347 & 2.694 \\
\hline
5 mm  & CoG               & 1.759 & 2.783 \\
      & Linear Regression & \textbf{1.419} & \textbf{1.990} \\
      & Random Forest     & 2.179 & 3.211 \\
      & MLP               & 1.721 & 2.259 \\
      & Waveform CNN      & 2.814 & 3.522 \\
\hline
\end{tabular}%
}
\end{table}

The results show that the best performing reconstruction method depends on the SiPM configuration. For the 20 mm configuration, CoG produced the lowest reconstruction error in both metrics, outperforming all the machine learning methods. For the \SI{10}{mm} pitch configuration, CoG achieved the smallest X residual width of \SI{1.134}{mm}, while Linear Regression achieved the lowest mean two-dimensional error of \SI{2.046}{mm}. The CoG result reported here for \SI{10}{mm} pitch is better than the value reported above. While for the CoG values reported in section \ref{sec:sipmReadoutGEM} an active area selection to the central area fully covered by SiPMs was used, the values reported in table \ref{tab:ml_reconstruction_results} use the same spatial subset of data used to evaluated the ML models. In the latter case, the CoG value reported was obtained from averaging results obtained in five separate test regions.

For the \SI{5}{mm} configuration, Linear Regression achieved the best performance in both metrics, with an X residual width of \SI{1.419}{mm} and a mean two-dimensional error of \SI{1.990}{mm}. Interestingly, the more complex neural network models did not consistently outperform Linear Regression, despite their ability to represent possible nonlinear relationships in the measured signals.

The lower reconstruction errors achieved by Linear Regression at 5 and \SI{10}{mm} pitch may be explained by the engineered waveform features and its ability to generalize to spatial regions not included in training. The normalized integrated signals, peak amplitudes, and peak positions provide additional information about the light-sharing pattern, allowing a simple model to capture useful relationships between the measured signals and particle-hit positions. Linear Regression learns a continuous mapping from these features to the reconstructed coordinates, which may help it generalize across the spatially separated test regions.

Overall, the results show that CoG remains a simple, robust, and physically interpretable method for reconstructing particle-hit positions. At the same time, machine learning methods show promising improvements in certain detector configurations. The results suggest that improving reconstruction accuracy does not necessarily require more complex models, but depends strongly on the detector configuration and how the measured signals are represented.

\section{Discussion}
\label{sec:5}

The achievable spatial resolution depends on sensor segmentation, light collection and photon statistics, as well as electron diffusion and the extent of the scintillation source. Where light is shared between neighbouring sensors, reconstruction using the measured light distribution can provide position estimates finer than the sensor pitch. This approach has been demonstrated for GEM scintillation read out with SiPM arrays \cite{Monteiro2024GEMSiPM}. The spatial granularity of such arrays is generally coarser than that of imaging cameras, but each channel can retain fast timing information, providing a useful combination of position sensitivity and temporal resolution.
In addition to spatial information, SiPMs can be used to obtain time information in optical TPCs. %Basic 3D track reconstruction combining the timing information from SiPMs with 2D images from a camera was demonstrated. Suitable SiPMs achieve sub-nanosecond intrinsic photon timing, although the resolution of a complete gaseous detector also depends on scintillation kinetics, electron transport, detected photon statistics and electronics \cite{Gundacker2020Timing}.
The small form factor, low cost and high sensitivity of SiPMs can be used to instrument optical TPCs with arrays of these photon detectors for resolving arrival time information of track segments with only a small influence on the optical readout by cameras. Thus, this combined readout approach could be used for more intricate track topologies, which cannot be easily resolved by the conventional combination of a camera with a single channel photon detector. 
These possibilities motivate SiPM readout of MPGDs as a compact and potentially simple detector architecture, whose performance must balance light collection, spatial sampling and channel count while accounting for dark counts, correlated noise and temperature-dependent gain.

\section{Conclusions}
\label{sec:6}

Optical readout with SiPMs is an alternative readout technique for MPGDs and can be used either stand-alone or in combination with camera-based readout. We reported on the readout of GEM and Micromegas detectors with an array of SiPMs and showed that millimetre-scale spatial resolution can be achieved with a centre-of-gravity reconstruction algorithm for events which emit sufficient scintillation light to be registered by multiple SiPMs.
While the placement of SiPMs below the amplification stage imposes some restrictions on the detector geometry, it seems preferential for equal signal sharing across the active area. We explored the alternative approach of placing SiPMs around the active area below a GEM stack but could not obtain reproducible signal sharing which would allow for successful center of gravity reconstruction.

%\acknowledgements
%\section*{Acknowledgements}
%\label{sec:8}

%\end{linenumbers}

\bibliographystyle{elsarticle-num-names} 
\bibliography{cas-refs}

%% else use the following coding to input the bibitems directly in the
%% TeX file.

\end{document}